\documentclass[aip,rsi,reprint,graphicx]{revtex4-1} 
\usepackage{graphicx}
\usepackage{hyperref}
\usepackage{upgreek}
\usepackage{natbib}

\draft 

\begin{document}


\title{A High-Performance, Low-Cost Laser Shutter using a Piezoelectric Cantilever Actuator} 



\author{W.~Bowden}
\affiliation{National Physical Laboratory, Teddington, Middlesex, UK }
\affiliation{University of Oxford, Oxford, Oxfordshire, UK }
\author{I.~R.~Hill}
\affiliation{National Physical Laboratory, Teddington, Middlesex, UK }
\author{P. E. G. ~Baird}
\affiliation{University of Oxford, Oxford, Oxfordshire, UK }
\author{P. ~Gill}
\affiliation{National Physical Laboratory, Teddington, Middlesex, UK }


\date{\today}

\begin{abstract}


We report the design and characterization of an optical shutter based on a piezoelectric cantilever.  Compared to conventional electro-magnetic shutters, the device is intrinsically low power and acoustically quiet.  The cantilever position is controlled by a high-voltage op-amp circuit for easy tuning of the range of travel, and mechanical slew rate, which enables a factor of 30 reduction in mechanical noise compared to a rapidly switched device.  We achieve shuttering rise and fall times of 11 $\upmu$s, corresponding to mechanical slew rates of $1.3 \textrm{ ms}^{-1}$, with an timing jitter of less than 1 $\upmu$s. When used to create optical pulses, we achieve minimum pulse durations of 250 $\upmu$s. The reliability of the shutter was investigated by operating continuously for one week at 10 Hz switching rate. After this period, neither the shutter delay or actuation speed had changed by a notable amount.  We also show that the high-voltage electronics can be easily configured as a versatile low-noise, high-bandwidth piezo driver, well-suited to applications in laser frequency control.

\end{abstract}


\maketitle 

\section{Introduction}

Optical shutters are ubiquitous in atomic physics and laser based experiments. The key figures of merit for evaluating the performance of these devices are: shutter speed, optical extinction and transmission, reproducibility of actuation delays, longevity of operation, and noise\textemdash both mechanical (acoustic, vibration) and electromagnetic. Furthermore, for transportable systems with the potential for space based applications, low power consumption and operation in zero gravity is necessary. Inevitability, cost considerations must also be addressed, particularly for complex experiments requiring many shutters.  

The majority of commercial and home built shutters use solenoid actuators \cite{Acharya} which are often relatively slow and noisey. With careful design these issues can be mitigated \cite{MaguireAndScholten,SingerAndWeidemüller},  however, such actuators require significant power to maintain shutter arm position and when actuated rapidly can broadcast electromagnetic noise to surrounding electronics. These drawbacks limit the utility of solenoid based shutters for many applications such as compact transportable experimental systems where electronics must be positioned in close proximity and consume little power. If sub-microsecond shutter speeds are required, acousto-optic\cite{Schwenger} and electro-optic devices are often used, although typically at the expense of increased insertion loss and reduced extinction. For our application in optical atomic clocks, incomplete extinction can lead to sizeable Stark shifts of the clock transition, effecting clock accuracy.  Although piezo based shutters have been demonstrated previously, they are typically discounted as useful optical shutters due to a limited actuation range leading to only partial extinction and transmission \cite{Adams,Aubin}. Our cantilever design leverages over a distance of a few centimetres the peizo's relatively small actuation range such that the shutter moves on the order of a millimetre\textemdash allowing it to extinguish comfortably or transmit a focused beam. After writing this paper, the authors became aware of a similar shutter design using cantilever piezos\cite{BauerAndWidera}. We build upon this work by presenting a complete mechanical and electrical design exhibiting improved performance.

\section{Shutter Design}

\subsection{Mechanical}

\begin{figure}[h]
    \centering
    \includegraphics[width=0.4\textwidth]{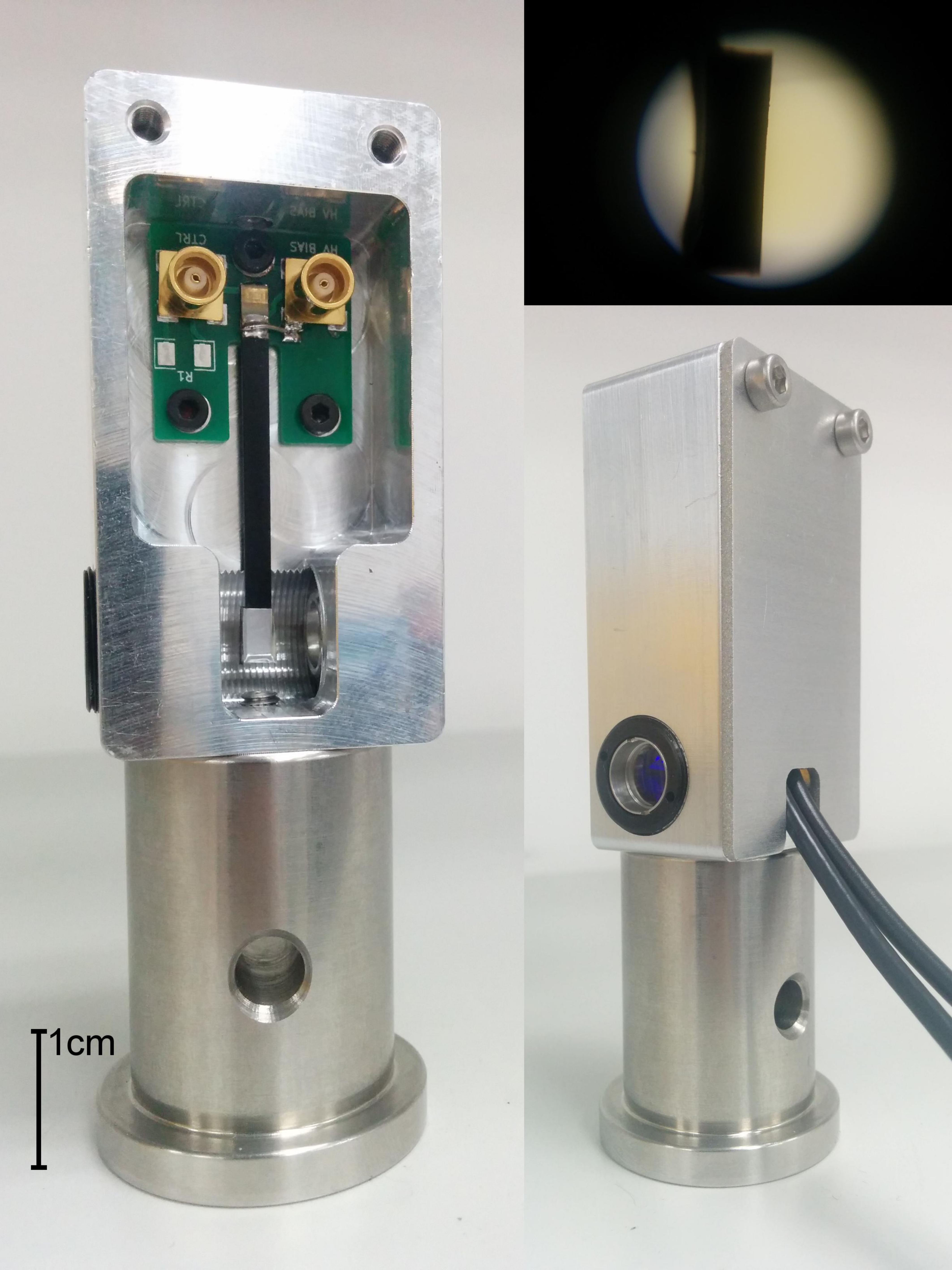}
    \caption{\label{fig:housing}The assembled piezo cantilever shutter. The shutter enclosure was designed to have a minimal footprint and offer various mounting configurations.  The shutter flag (top right) can be imaged using the input lens to ensure it is cleanly cut without any debris.}
\end{figure}

The shutter, shown in \autoref{fig:housing}, consists of a thin black-coated foil flag glued to the tip of a cantilever piezoelectric actuator\footnote{APC International Ltd., formerly American Piezo Ceramics Inc., product description 350/025/0.70SA.} positioned at the focal point between two aspheric lenses.  The piezoelectric cantilever is either soldered, or silver epoxy bonded, to a small PCB which is secured inside an aluminum enclosure measuring $2\times3\times5$ cm$^3$. The enclosure has two threaded apertures to hold two aspheric lenses for 1) focusing the input beam to the shutter flag and 2) re-collimation on exit. In our application the shutters are typically placed immediately before optical fibres and provide additional freedom via the re-collimation lens to optimise fibre input-coupling efficiency. In operation, the piezo receives a high voltage bias (150V) and control voltage (0 to 150V) through two MCX connectors. Care should be taken when soldering the cantilever to prevent thermal damage of the piezo ceramic.

\subsection{Electronics}
\label{electronics}

\begin{figure}[h]
    \centering
    \includegraphics[width=0.45\textwidth]{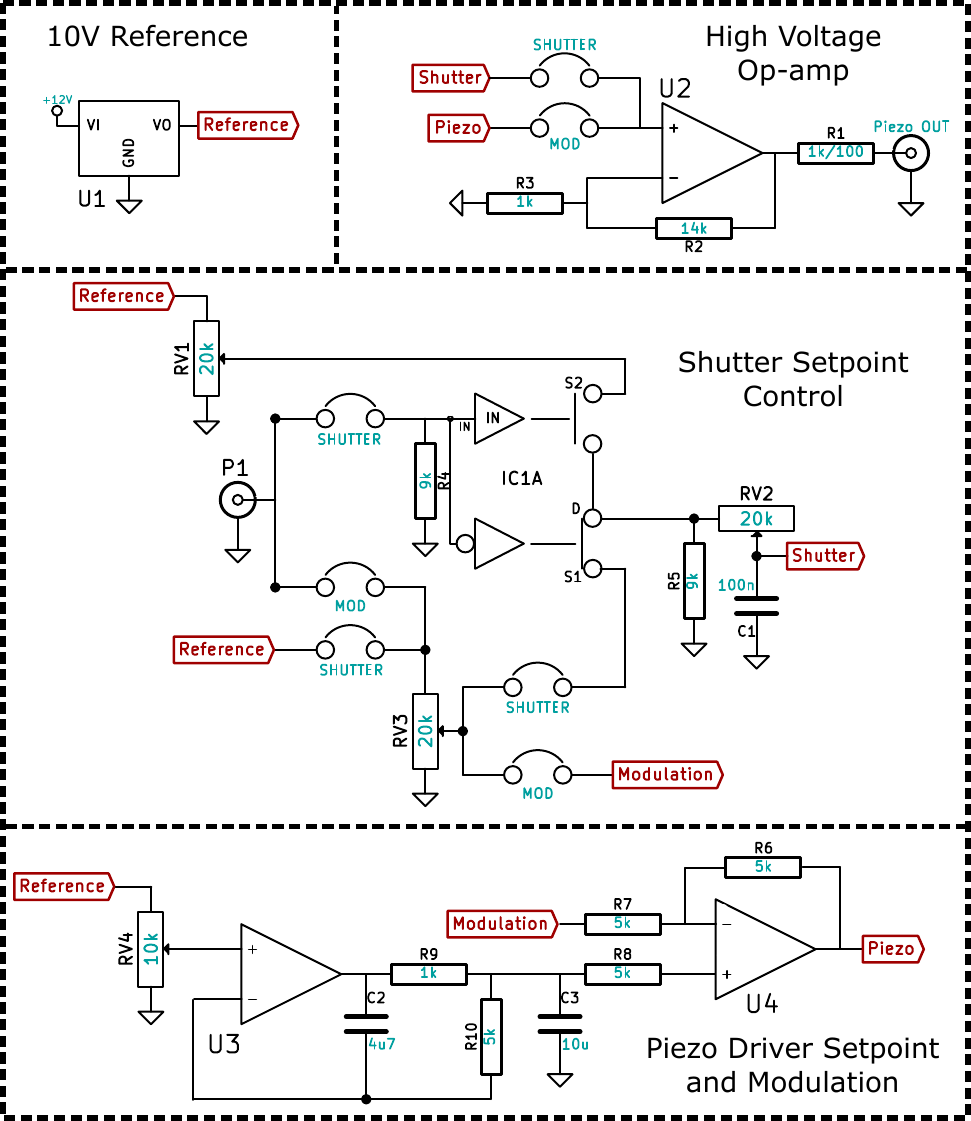}
    \caption{Schematic of the high voltage driver used to power the shutters. By switching jumper pairs (labelled `MOD' and `SHUTTER'), the circuit can be configured to act as either a shutter driver or as a general purpose low-noise piezo driver. For shutter operation, the shutter is toggled by applying a TTL signal to `P1', with the open and closed positions set by `RV1' and `RV3'. The slew rate is controlled by `RV2'. For operation as a general purpose piezo driver, the high voltage output level is set by `RV4' and can be modulated using the input `P1' with a sensitivity controlled by `RV3'. In this application, the output impedance `R1' is typically lowered from 1 k$\Omega$ to 100 $\Omega$. `U1' is a 10V precision voltage reference (LT1021-10V), `U2' is a high voltage op-amp (PA441), `U2' and `U3' are single supply low noise op-amps (LT1006), and `IC1' is a TTL controlled analogue switch (ADG419). Not shown are voltage regulators for the $\pm$12V rails and any decoupling or compensation capacitors for the op-amps or voltage references. These values can be found in their respective data sheets. Also not shown is the high voltage supply needed to power `U2' and bias the piezo.}
    \label{fig:board}
\end{figure}

The control circuit, shown in \autoref{fig:board}, uses a high-voltage op-amp\footnote{Apex Instruments PA441. The APEX PA341 device was also tested and while adequate for shuttering applications, the broadband noise was over 10 times greater and unsuitable for low-noise piezo driving applications.} (U2) powered by a high-voltage DC-DC converter\footnote{American Power Design, part A15.} (not shown). Given the low power consumption of the device, we can power up to 12 shutter boards from a single high-voltage supply. A TTL controlled analogue switch (IC1A) selects for amplification by U2 one of two tuneable control voltages (RV1 and RV3) to set the open and close positions of shutter. This allows convenient tuning of the cantilever position and range of travel. Prior to be being amplified, these control voltages are low-pass filtered to effect a reduction in the mechanical slew rate of the shutter arm (determined by RV2 and C1). By limiting the range of travel and mechanical slew rate of the cantilever, we reduce vibrations, decrease shuttering delays, and through reducing stress on the piezo postulate an improvement in lifetime of the device.  To increase utility of the control electronics the circuit can be easily reconfigured using on-board jumpers to serve as a tunable piezo driver for use in, for example, laser frequency stabilization applications\textemdash where both easy tuning and fast modulation of the output voltage, together with low voltage noise, are required. In this configuration, the output voltage is set using a trim-pot which divides down a stable voltage reference (U1)\textemdash the same reference is used to derive the two shutter set points. This voltage is filtered by an actively compensated low-pass filter (using U3) with pole determined predominantly by R9 and C3 ($f_{-3\mathrm{dB}} = 1/2\pi~\mathrm{R9}\cdot \mathrm{C3}$), with R10 and C2 required for stability (satisfying $\mathrm{R10}\cdot \mathrm{C2} > 2\,\mathrm{R9}\cdot \mathrm{C3}$) before being combined in U4 with an external control voltage at P1. This control voltage allows for fast modulation of the output voltage, limited by the power available to the high-voltage circuit. \footnote{For more details regarding the design of the shutter including  bill of materials and machinable drawings, please contact the author.}

%
%

\section{Performance}

\subsection{Shutter Speed and Extinction}

\begin{figure}[h]
    \centering
    \includegraphics[width=0.49\textwidth]{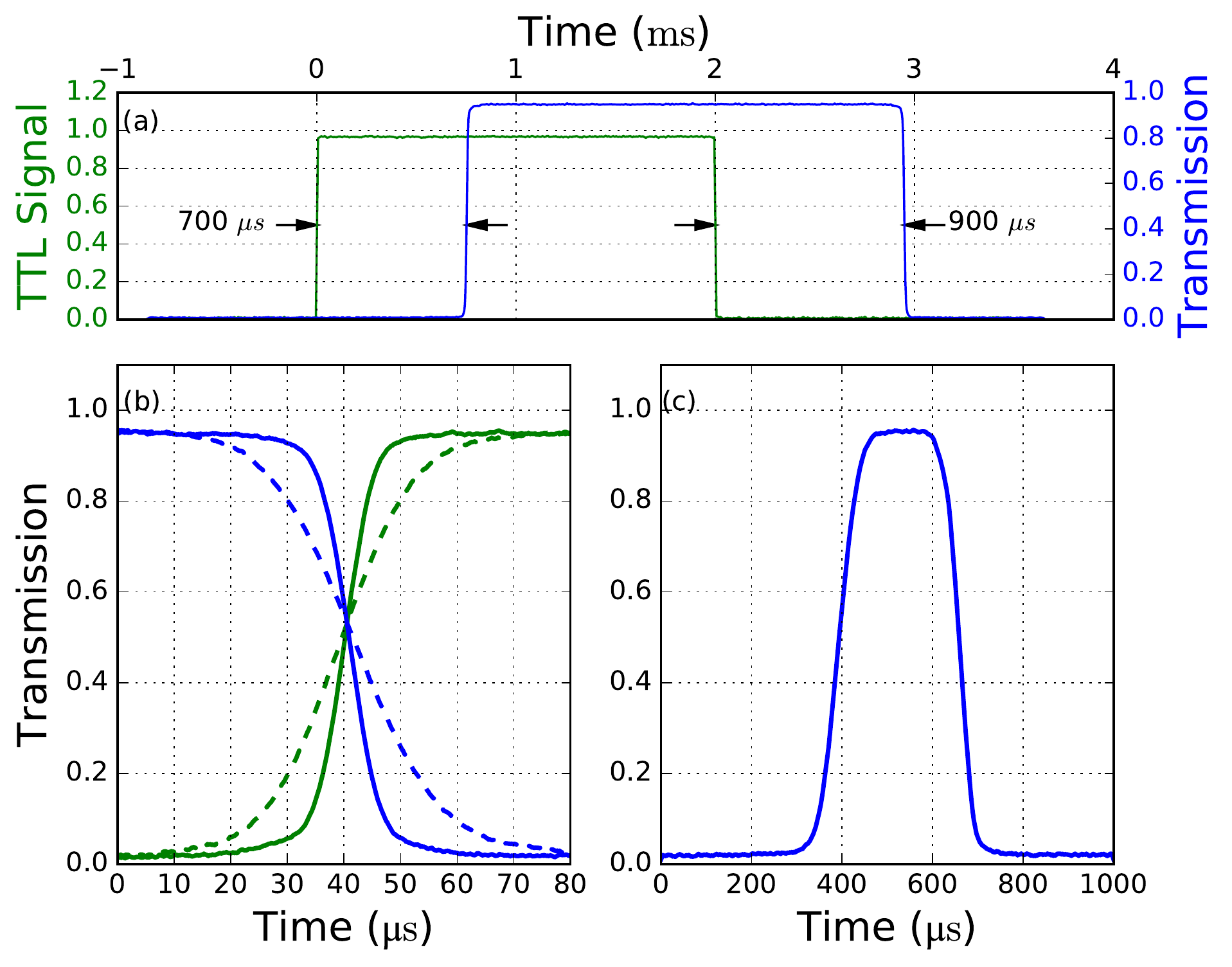}
    \caption{(a) The response of the shutter to a TTL control signal at time t=0~ms and t=2~ms.  The shutter `open' and `close' delays are 0.7~ms and 0.9~ms, respectively.  (b) A photo-diode signal showing the rise and fall times, as defined by the 10$\%$\textendash90$\%$ transmission crossing points, for the shutter with the fastest (solid line) and slowest (dotted line) slew rates, set by `RV2' (\autoref{fig:board}). In the slowest case, with the lowest vibration, the switching time is approximately 30~$\upmu$s while for the faster setting it is 11~$\upmu$s. This corresponds to mechanical slew rates of between 0.5-1.3~ms$^{-1}$.  (c) The profile of a sub-millisecond pulse produced by the shutter.}
    \label{fig:speed}
\end{figure}

To test the shutter, a collimated Gaussian beam from a $655$ nm laser was focused to an 11 $\upmu$m $1/e^{2}$ beam waist by the input lens\footnote{A220TM-B - f = 11.0 mm, NA = 0.26, Mounted Rochester Aspheric Lens} and aligned on to the shutter flag. The piezo cantilever has a manufacturer's quoted mechanical resonance at approximately 200 Hz which is excited in rapid shutter actuation leading to damped oscillations, or a settling period, of the cantilever. As a result, the `closed' and `open' positions must be set slightly beyond the minimum required distance to prevent these excursions from modulating the optical transmission as the cantilever settles. Reducing the mechanical slew rate helps to suppress these oscillations, but this comes at the expense of switching speed. \autoref{fig:speed} shows the actuation of the shutter for the slowest and fastest shuttering speeds, corresponding to rise and fall times (measured period between $10\%\textendash90\%$ transmission points) of \mbox{11\textendash33~$\upmu$s}. Given the beam size, this corresponds to mechanical slew rates between 0.5\textendash1.3~ms$^{-1}$ making it comparable to solenoid based shutters \cite{MaguireAndScholten,Acharya}. The transmission is limited to approximately $95\%$ by the anti-reflection coating of the two lenses while the extinction when closed is greater than 100~dB.  An exact measurement of the extinction was limited by the noise floor of the photodiode used in the test setup. Additionally, the shutter can be opened momentarily to produce short optical pulses with durations as short as 250 $\upmu$s. The shutter can be easily modified to produce much shorter pulses by cutting a V-shaped notch in the shutter flag \cite{Scholten} such that the beam is both transmitted and extinguished during a single pass. In such a scheme the pulse duration is easily controlled by tuning the mechanical slew rate.

\subsection{Durability and Reproducibility}

To test the durability of the shutter we operated a device at a 10 Hz switching rate for one week (6 million cycles). After this period the shuttering speed and delay had not differed from the values measured prior to testing. Recently we operated 10 shutters consistently for three weeks at a constant switching rate of 2 Hz. Again, we observed no change in shutter behaviour throughout this extended period of operation.

\begin{figure}[h]
    \centering
    \includegraphics[width=0.49\textwidth]{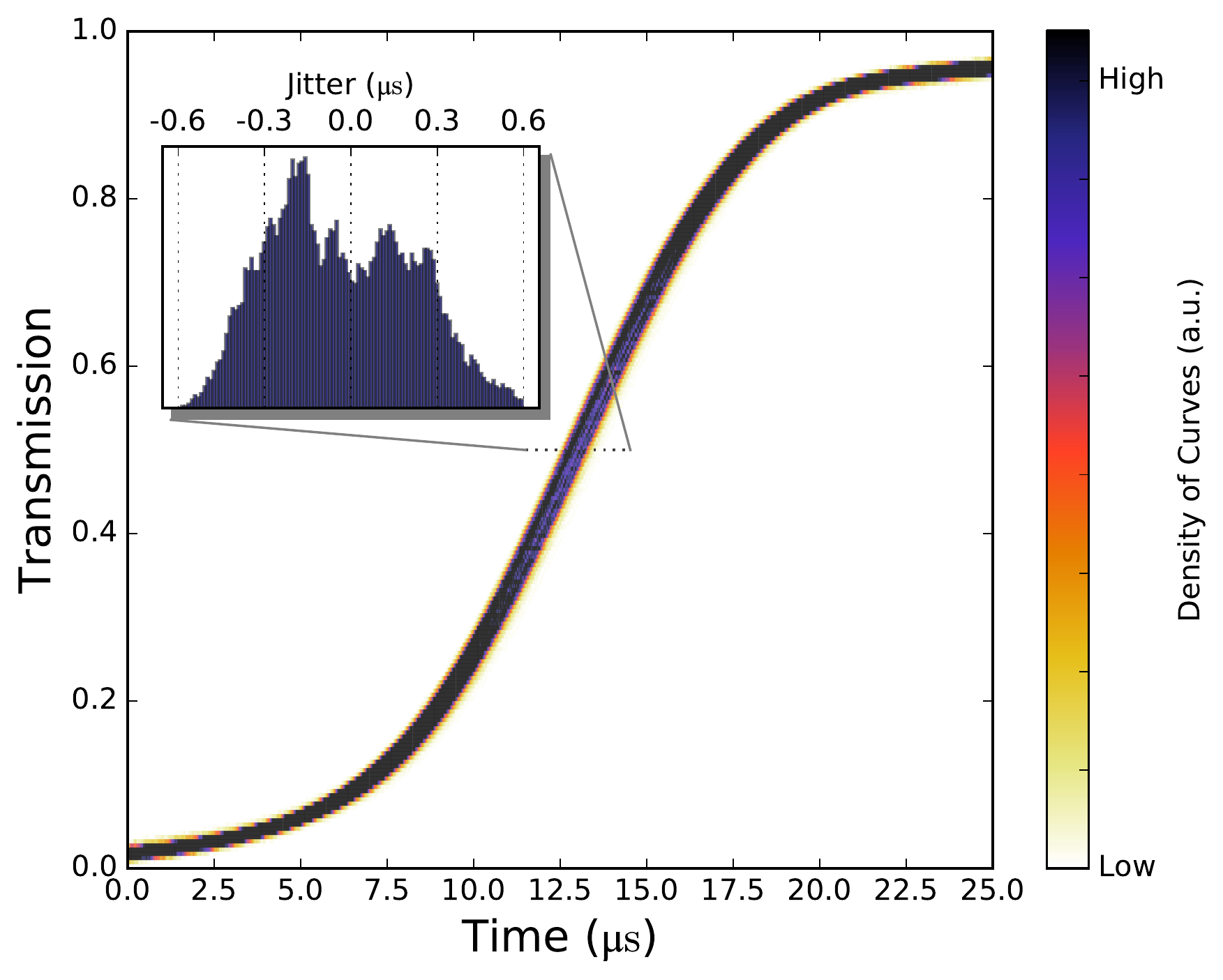}
    \caption{An eye-plot showing transmission of the test beam during transit of the opening shutter for one thousand cycles. A histogram of the time to reach 50\% transmission, with mean removed, is shown in the inset. The timing jitter is less than 1 $\upmu$s.}
    \label{fig:eyePlot}
\end{figure}

Variation in the shot-to-shot transit of the shutter referenced to the TTL control pulse was measured for 1000 switching cycles. \autoref{fig:eyePlot} shows the timing jitter in the 50\% transmission point to be less than 1 $\upmu$s.  The typical actuation delay is between 0.75~ms and \mbox{1.25~ms} and is tuned by setting the limits of the piezo actuation range and slew rate.

\subsection{Acoustic Noise and Vibration}

\begin{figure}[h]
    \centering
    \includegraphics[width=0.49\textwidth]{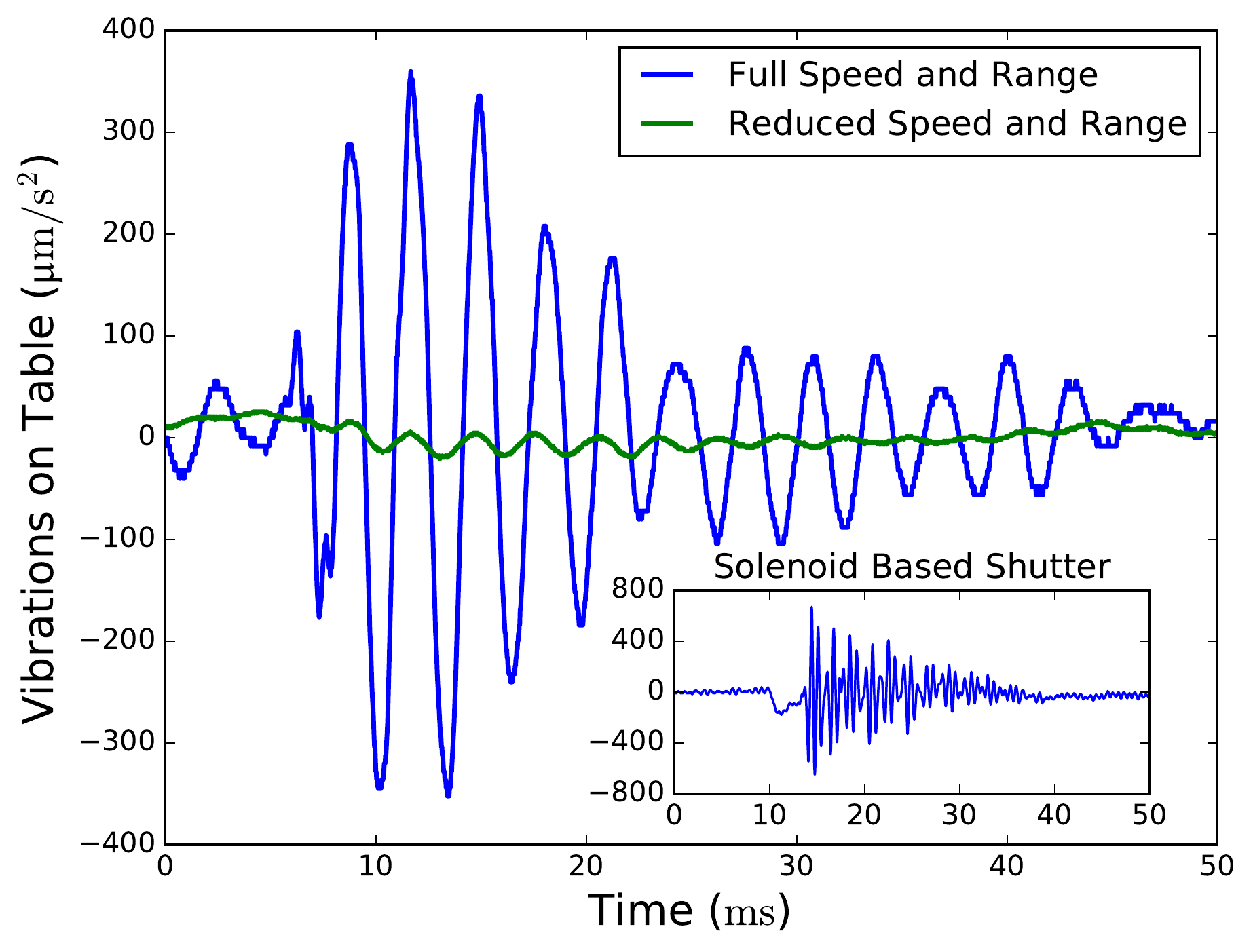}
    \caption{Mechanical vibrations measured in the immediate surroundings of the shutter rigidly mounted to an optical table. A factor of 30 reduction is achieved by limiting the range of travel and mechanical slew rate of the actuator. Slowing the shutter and limiting its range resulted in approximately equal parts reduction in vibration. For comparison, the measured vibration of a solenoid based shutter is shown in the inset.}
    \label{fig:vibrations}
\end{figure}

An accelerometer was used to measure vibrations caused by the shutter actuation, see \autoref{fig:vibrations}. In comparison to operating the shutter without any compensation, our piezo driver board can reduce the vibrations by a factor of 30 by limiting the range of travel and mechanical slew rate.  When properly configured, the piezo cantilever based design generates significantly less vibration in comparison to our previously used solenoid based shutter, see inset \autoref{fig:vibrations}. The shutter is acoustically quiet.

\subsection{Piezo Driver Electrical Noise}

\begin{figure}[h]
    \centering
    \includegraphics[width=0.45\textwidth]{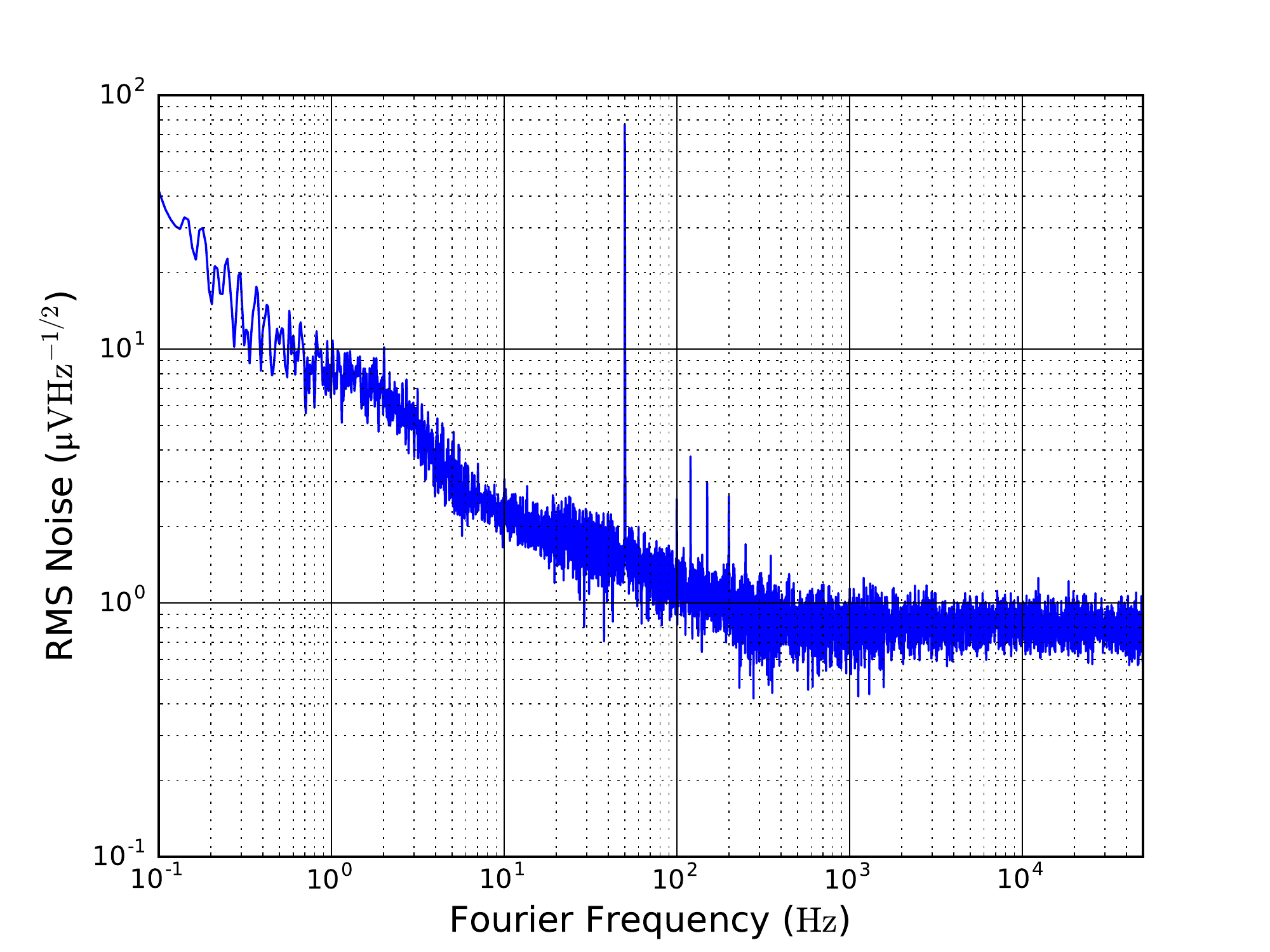}
    \caption{Power spectral density of the noise on the output of the high voltage driver exhibiting a noise floor below \mbox{1 $\upmu \textrm{V}\textrm{Hz}^{-1/2}$} (1 M$\Omega$ load).}
    \label{fig:noise}
\end{figure}

As described in section \ref{electronics}, the control circuit is easily configured for use as a general purpose piezo driver with high modulation bandwidth, making it ideal for controlling laser cavities or tunable Fabry-Perot resonators. In such applications it is desirable for the driver to exhibit low voltage noise. For this configuration, the voltage noise power spectral density of the high-voltage output is shown in \autoref{fig:noise}. The low noise properties are intrinsic to the high voltage op-amp and are not a result of any excessive filtering which can limit the modulation bandwidth of the driver. The small signal modulation bandwidth of the driver was measured to be 160 kHz (\mbox{$-3$~dB} point for a 1~V$_{\mathrm{pp}}$ output modulation with 1~M$\Omega$ load). When driving a piezo, a pole is added with a time constant set by the capacitance of the piezo and the output impedance of the circuit. For example, with an output independence of 100 $\Omega$ and piezo capacitance of 10 nF, the $-3$~dB point dropped to approximately 100~kHz. If such a piezo driver were to be used in open loop to actuate the length of an external cavity diode laser, the contribution of this voltage noise to the laser's spectral newline can be estimated using the relation between the frequency noise spectrum and the laser line shape\cite{DiDomenico:10}. Assuming a frequency tuning sensitivity of 50 MHzV$^{-1}$ for the laser, the contribution of this voltage noise to the FWHM  linewidth with an observation time of one second would be approximately 7~kHz. In reality this is an upper bound on the estimated broadening as the finite mechanical bandwidth of the piezo-actuated retro reflector filters the transfer of voltage noise to frequency excursions.

\section{Discussion}

We have demonstrated a compact, high-performance optical shutter based on a cantilever piezoelectric actuator. Our characterisation supports these claims showing fast shuttering with minimal timing jitter and acoustic noise for long periods of operation. Previously, such a piezo based shutter solution may have appeared unattractive due to the requirement of switching high voltages. However, with the current availability of high-performance, high-voltage, low-noise op-amps, the task of circuit design is a less onerous one. Furthermore, we showed how the control circuit can be easily modified for use as a general purpose piezo driver with noise properties rivaling those of commercial offerings and with extended modulation bandwidth.

\section{Acknowledgments} 

The authors give acknowledgment to Peter Nisbet-Jones and Steven King for the initial work investigating piezo cantilever shutters and helping with the vibration noise measurements. Acknowledgment also goes to Richard Hobson for help with shutter testing and Charles Baynham for helpful suggestions regarding the manuscript. This work was done under the auspices of both the Future Atomic Clock Technology Initial Training Network funded by the European Union and the UK NMO program.

\bibliography{myBib}

\end{document}